\documentclass[preprint]{aastex}

\tighten
\input psfig.tex

\slugcomment{submitted to ApJ.}
\newcommand{\mpc}{h^{-1}\ {\rm Mpc}}
\newcommand{\etal}{{\it et al.\ }}
\newcommand{\xiav}{\bar{\xi}}

\newcommand{\avg}[1]{\langle{#1}\rangle}

\newcommand{\ltsima}{$\; \buildrel < \over \sim \;$}
\newcommand{\lsim}{\lower.5ex\hbox{\ltsima}}
\newcommand{\gtsima}{$\; \buildrel > \over \sim \;$}
\newcommand{\gsim}{\lower.5ex\hbox{\gtsima}}
\newcommand{\eg}{{\it e.g.,\ }}
\newcommand{\ie}{{\it i.e.\ }}

\slugcomment{Submitted to {\it The Astrophysical Journal}, August 2000}
\shortauthors{Szapudi et al.}
\shorttitle{Higher Order Clustering up to $z\simeq 1$}

\begin{document}

\title{Observational Constraints on Higher Order Clustering up to $z\simeq 1$}

\author{Istv\'an Szapudi}
\affil{CITA, University of Toronto, 60 St George St\altaffilmark{1},
Toronto, Ontario, M5S 3H8}

\author{Marc Postman\altaffilmark{2}}
\affil{Space Telescope Science Institute\altaffilmark{3},
Baltimore, MD 21218}

\author{Tod R. Lauer}
\affil{National Optical Astronomy Observatories\altaffilmark{4},
Tucson, AZ 85726}

\author{William Oegerle\altaffilmark{2}}
\affil{Johns Hopkins University, Department of Physics \& Astronomy,
Baltimore, MD 21218}

\vfill

\altaffiltext{1}{E-mail:\ szapudi@cita.utoronto.ca}
\altaffiltext{2}{Visiting Astronomer Kitt Peak National Observatory,
NOAO.}
\altaffiltext{3}{The Space Telescope Science Institute
is operated by the Association of Universities for
Research in Astronomy (AURA), Inc., under National Aeronautics and
Space Administration (NASA) Contract NAS 5-26555.}
\altaffiltext{4}{The National Optical Astronomy Observatories are
operated by AURA, Inc., under cooperative agreement with the National
Science Foundation.}       

\begin{abstract}

Constraints on the validity of the hierarchical gravitational
instability theory and the evolution of biasing are presented based upon 
measurements of higher order clustering statistics in the Deeprange
Survey,  a catalog of $\sim710,000$ galaxies with $I_{AB} \le 24$
derived from a KPNO 4m CCD imaging survey of a contiguous
$4^{\circ} \times 4^{\circ}$ region. 
We compute the 3--point and 4--point angular correlation functions
using a direct estimation for the 
former and the counts-in-cells technique for both. 
The skewness $s_3$ decreases by a factor of $\simeq 3-4$
as galaxy magnitude increases
over the range $17 \le I \le 22.5$ ($0.1 \lesssim z \lesssim 0.8$).
This decrease is consistent with a small {\it increase} of the bias
with increasing redshift, but not by more than a factor of 2 for the highest
redshifts  probed. Our results are strongly inconsistent, at about the
$3.5-4\ \sigma$ level, with typical cosmic string models in which the 
initial perturbations  follow a non-Gaussian 
distribution -- such models generally predict an opposite trend in the
degree of bias as a function of redshift.
We also find that the scaling relation between the 3--point and 4--point
correlation functions remains approximately invariant over the above magnitude
range. The simplest model that is consistent with these constraints is
a universe in which an initially Gaussian perturbation spectrum evolves 
under the influence of gravity combined with a low level
of bias between the matter and the galaxies that decreases slightly from
$z \sim 0.8$ to the current epoch.

\end{abstract}


\section{Introduction} 

The evolution of the spatial distribution of galaxies is intimately 
related to the physical processes of galaxy formation, to the initial
spectrum and subsequent gravitational growth of matter fluctuations 
in the early universe,
and to the global geometry of space-time. Quantifying the galaxy
distribution is, thus, fundamental to cosmology and has dominated
extragalactic astronomy for the past two decades. The $n-$point
correlation functions provide a statistical toolkit that
can be used to characterize the distribution.

The two-point correlation function is the most widely used statistic
because it provides the most basic measure of galaxy
clustering -- the departure from a pure Poisson distribution. 
It is also popular because its execution is computationally straight forward. 
The two-point correlation function is defined as the joint
moment of the galaxy fluctuation field, $\delta_g$, at two different positions
\begin{equation}
\xi_2 = \xi = \avg{\delta_{g,1}\delta_{g,2}} \label{eq:xidef},
\end{equation}
where $\avg{}$ means ensemble average. 
The two-point correlation function yields a full description
of a Gaussian distribution only, for which all higher order
connected moments are zero by definition.
The galaxy distribution, however, exhibits non-Gaussian
behavior on small scales due to non-linear gravitational amplification of
mass fluctuations, even if they grew from an initially Gaussian field.
On larger scales, where the density field is well represented by linear
perturbation theory, non-Gaussian behavior may still be present if
the initial perturbation spectrum was similarly non-Gaussian. 
In addition, the process of galaxy formation is likely to introduce
biases between the clustering properties of the dark and luminous matter.
In the presence of such realities, higher order moments are required to
obtain a full statistical description of the galaxy distribution
and to provide discrimination between different biasing scenarios
(\eg, Fry \& Gazta\~naga 1993, Fry 1994, Jing 1997).

An accurate determination of higher order clustering statistics requires 
a large number of galaxies and, to date, the most accurate measurements
have been derived from wide-area angular surveys, 
such as the APM (Szapudi \etal 1995; Gazta\~naga 1994)
and the EDSGC (Szapudi, Meiksin, \& Nichol 1996), 
although recent redshift surveys are
now becoming large enough to make interesting constraints
(e.g., Hoyle, Szapudi, \& Baugh 1999; Szapudi \etal 2000a). 
These surveys are, by design, limited to the
study of the current epoch galaxy distribution. 
Deep surveys add a further dimension to the exploration
of clustering by enabling the study of its evolution.
Ideally one would like to have deep, wide redshift surveys
available for such analyses but it's observationally infeasible at the
current time. Projected surveys thus still provide a unique way
to study the evolution of clustering, especially if photometric redshifts
are available. Several deep surveys have been used to study 
the evolution of low order galaxy clustering 
(Lilly \etal 1995; Le Fevre \etal 1995; Neuschaffer \& Windhorst 1995;
Campos \etal 1995; Connolly \etal 1996; Lidman \& Peterson 1996;
Woods \& Fahlman 1997; Connolly, Szalay, \& Brunner 1998)
but most suffer from insufficient contiguous area and are therefore
barely large enough to measure the two-point correlation function. 
The Deeprange survey (Postman \etal 1998; hereafter paper I) 
was designed to study the 
evolution of clustering out $z \sim 1$. The resulting catalog contains
$\sim710,000$ galaxies with $I_{AB} \le 24$
derived from a KPNO 4m CCD imaging survey of a contiguous
$4^{\circ} \times 4^{\circ}$ region. The photometric calibration of the
catalog is precise enough over the entire survey to 
limit zeropoint drifts to $\lsim 0.04$ mag 
that translates to a systematic error in the angular two-point
correlation function, $\omega(\theta)$, of $\lsim 0.003$
on a $4^{\circ}$ scale and proportionally less on smaller scales 
(see paper I for details). Accurate measurements of the $I$-band number
counts over the range $12 < I_{AB} < 24$ and the two--point angular correlation
function up to degree scales are presented in paper I. 

The size and quality of the Deeprange catalog
are sufficient to enable reliable estimation of the 3--point and 4--point 
angular correlation functions as well.
This paper presents the first attempt to constrain higher order 
correlation functions down to flux limits of $I_{AB} = 23$, corresponding
to an effective redshift of $z \simeq 0.75$.  We briefly review the
astrophysics contained within the higher moments in section \ref{sec:statrev}.
Section \ref{sec:anal} presents a description of the galaxy sample analyzed and
the computational methods
used to derive the statistics and section \ref{sec:results} summarizes the
results. The implications of our calculations are presented in 
section \ref{sec:sum}. Technical issues concerning the fit
of the three-point correlation are presented in Appendix A.

\section{A Brief Review of Higher Order Statistics}
\label{sec:statrev}

We begin with a summary of the basic definitions of the higher order
statistical methods used in this paper, highlighting their
most fundamental properties. For a more in depth review, the 
reader should consult the references cited.

The observed large-scale structures in the local universe are
characterized by a high degree of coherence (e.g. de Lapperent, Geller, \& 
Huchra 1986; Shectman et al. 1996; da Costa et al. 1998) 
and some features, like the ``Great Wall" have undergone asymmetric 
gravitational collapse (dell'Antonio, Geller, \& Bothun 1996).
Furthermore, the galaxy distribution traces the 
underlying dark matter in a non-linear way that may depend on time and scale. 
It is the non-zero higher order correlation functions that
uniquely characterize such phenomena and allow discrimination between
the observed galaxy distribution and a Gaussian distribution with the same 
variance,
\ie two-point correlation function. The $N$-point correlation function 
\begin{equation}
\xi_N = \avg{\delta_1\delta_2\ldots\delta_N}
\end{equation}
depends on a large number of parameters 
($3N$ coordinates, minus 3 rotations and 3 translations). 
The number of parameters decreases
if the function is integrated over part of the configuration space
(the geometric distribution of the $N$-points is often referred to as their
{\it configuration}, and the $N$-dimensional space describing it as the
configuration space). 

A simple model for the $N$-point correlation functions is the
clustering hierarchy (e.g., Peebles 1980)
defined as
\begin{equation}
   \xi_N({r_1},\ldots, {r_N}) = \sum_{k=1}^{K(N)}
   Q_{Nk} \sum^{B_{Nk}} \prod ^{N-1} \xi(r_{ij}),
   \label{hierar}
\end{equation}
where $\xi(r) \equiv \xi_2(r) = (r/r_0)^\gamma$, 
and $Q_{Nk}$ are structure constants. Their average is
\begin{equation}
Q_N = \frac{ \displaystyle {\sum_{k=1}^{K(N)} Q_{Nk}B_{Nk} F_{Nk}}}
          { \displaystyle {N^{(N-2)}}},
\end{equation}
where $F_{Nk}$ are the form factors associated with the shape
of a cell of size unity (see Boschan, Szapudi \& Szalay 1994 for details).
% The product above runs over the $N-1$ edges of a tree.
% The summation in equation~(\ref{hierar}) is over
% all possible $N^{N-2}$ trees with $N$ vertices.
% In the sum, every $\xi(r_{ij})$ corresponds to an edge
% $r_{ij}=\abs{{r_i-r_j}}$ in a tree spanned by ${r_1}, \ldots,
% {r_N}$. For each tree, there is a product of $N-1$ two-point
% functions and a summation over all the $B_{Nk}$ labelings of
% all the $K(N)$ distinct trees. 
For the three-point correlation function, our main concern here, 
the above form factors amount to a few percent only; if neglected 
then $Q_3 \simeq S_3/3$ (see the definition of $S_3$ below).

If the integration domain is a particular cell of volume $v$,
with the notation $\bar f = \int_v f/v$ for cell averaging,
then the amplitude of the $N$-point correlation
function can be expressed as 
\begin{equation}
  S_N = \bar\xi_N/\bar\xi^{N-1} = \avg{\delta^N}_c/\avg{\delta^2}^{N-1}.
\end{equation}
The $S_N$'s, with a suitable normalization motivated by leading
order perturbation theory, are commonly used 
to characterize the higher order, i.e.
non-Gaussian, properties of the galaxy distribution in real surveys
and N-body simulations.
In the second half of the above equation the integration over
the cell, i.e. smoothing (or filtering) is implicit. While
the $S_N$'s do not retain all the information encoded in the
$N$-point correlation functions, in particular their shape dependence, 
it is an extremely useful measure of clustering. It is directly
related to the distribution of counts in cells (in the same
cell $v$), as it is the cumulant or connected moment
\footnote{Ensemble averages over the connected component have the
statistical property of additivity for independent
processes and hence receive the name ``cumulants''. The alternative
terminology ``connected'' comes from the (Feynman) graph
representation of the statistical processes. It can
be shown that cumulants correspond to graphs that
have exactly one connected component. The mathematical
definition relies on a logarithmic mapping of the
generating function of ordinary moments (e.g. Szapudi
\& Szalay 1993)} thereof.
If the shape of the cell $v$ is fixed, these quantities depend
only on one parameter, the size of the cell.
Note the alternative notation $Q_N = S_N/N^{N-2}$. These
two notations differ only in their normalizations:
for $Q_N$ it follows from the hierarchical assumption
(see later),  and for $S_N$ from perturbation theory.
For Gaussian initial conditions with power spectrum
of slope $n$,
leading order (tree-level) perturbation theory of the underlying 
density field in an
expanding universe predicts the $S_N$'s. For $N = 3$ (skewness)
and $N = 4$ (kurtosis) the prediction is 
(Peebles 1980; Fry 1984; Juszkiewicz, Bouchet, \& Colombi 1993;
Bernardeau 1994; Bernardeau 1996):
\begin{eqnarray}
  S_3 && = \frac{34}{7} - (n+3)\cr
  S_4 && = \frac{60712}{1323} - \frac{62(n+3)}{3} +
        \frac{7(n+3)^2}{3}.
\end{eqnarray}
These equations depend on scale through the local slope, $n$,
of the initial power spectrum that, except for
scale invariant initial conditions, varies slowly with
scale in all popular cosmological models. According to
perturbation theory and simulations, 
geometric corrections from $\Omega, \Lambda$ are negligible,
(e.g., Bouchet \etal 1995).
The above results are valid on scales $\gtrsim 7\mpc$. On smaller
scales accurate measurements from $N$-body simulations exist
(e.g., Colombi, Szapudi, Jenkins, \& Colberg 2000, and references
therein) but the measurements are subject to a particular cosmological
model.

If the galaxy field, $\delta_g$, is a general non-linear function of
the mass density field, $\delta$, the we can express this function
as a Taylor series $f(\delta)  = b_1\delta+b_2 \delta^2/2 + \ldots$. 
The $b_N$ are the non-linear biasing coefficients, of which
$b_1 = b$ is the usual bias factor connecting the two-point correlation
function of galaxies with that of the dark matter as
\begin{equation}
  \xi_g = b^2 \xi.
\end{equation}
For higher order correlation functions, and for the $S_N$'s,
analogous calculations relate the statistics of the galaxy
and dark matter density fields 
(Kaiser 1984; Bardeen \etal 1986; Grinstein \& Wise 1986;
Matarrese, Lucchin, \& Bonometto 1986; Szalay 1988; Szapudi 1994;
Matsubara 1995; Fry 1996; Szapudi 1999).
For example, the result for third moment is (Fry \& Gazta\~naga 1993)
\begin{equation}
  S_{3,g} = \frac{S_3}{b} + \frac{3 b_2}{b^2}.
\end{equation}
This formula is expected to hold on the same or larger scales 
as leading order (weakly non-linear) perturbation theory, 
i.e. on scales $\gtrsim 10\mpc$.
On smaller scales, 
Szapudi, Colombi, Cole, Frenk, \& Hatton  (2000) found the
following phenomenological rule from N-body simulations 
\begin{equation}
  S_{N,g} = \frac{S_N}{b_*^{2(N-2)}},
  \label{eq:sqbias}
\end{equation}
where the function $b_* = b_*(b) \simeq b$.
For $b \gtrsim 1$ the typical effect of biasing is that it
decreases the $S_N$'s. The above theory does not include
stochastic effects, when the galaxy density field is a 
random function of the underlying dark matter field.
Stochasticity typically introduces only a slight extra
variance on the parameters of the theory and, thus,
it will not be considered further in this paper.

\section{Sample Definition and Analysis Methods}
\label{sec:anal}

The construction of the galaxy catalog is described in paper I. Briefly,
the $I$-band survey consists of 256 overlapping CCD images. The field of view 
of each CCD is 16 arcminutes but the centers are spaced 15 arcminutes apart. 
The 1 arcminute overlap enables accurate astrometric and photometric 
calibration to be established over the entire survey. Objects were identified, 
photometered, and classified using automated software. For consistency, we 
analyze the higher order clustering properties in the same magnitude 
slices\footnote{Magnitude ranges are given in the Kron-Cousins $I-$band;
the conversion to AB magnitudes is achieved by adding $\sim0.5$ mag
to the Kron-Cousins values.} 
used to measure the two--point correlation functions in paper I.

\subsection{Counts in Cells}

As our goal is to describe the higher order clustering of galaxies,
we adopt statistical measures that are closely
related to the moments of the underlying density field. 
The estimation of the higher order correlation amplitudes
follows closely the method described in Szapudi, Meiksin, \& Nichol (1996), 
which can be consulted for more details. Only the most relevant
definitions are given next, together with the outline of the
technique.

The probability distribution of counts in cells, $P_N(\theta)$,
is the probability that an angular cell of dimension $\theta$ contains $N$ 
galaxies.  The factorial moments of this distribution,
$F_k = \sum P_N (N)_k$ (where $(N)_k = N(N-1)..(N-k+1)$ is
the $k$-th falling factorial of $N$),
are indeed closely related to the moments of the underlying 
density field, $\avg{N}(1+\delta)$, through
$\avg{(1+\delta)^k} = F_k/\avg{N}^k$ (Szapudi \& Szalay 1993).

The most common method employed to relate the discrete nature
of the observed galaxy distribution to the underlying continuous density field
is known as infinitesimal Poisson sampling, which is effectively a
shot noise subtraction technique. This corresponds to the assumption
that, for an infinitesimal cell, the number of galaxies follows a Poisson
distribution with the mean determined by the underlying field.
This assumption must be approximate
for galaxies, especially on small scales, because of
possible halo interaction or overlap. Nevertheless, on the
scales we are studying, Poisson sampling should be a good approximation.
Moreover, even on scales where no underlying continuous process exists,
factorial moments are the preferred way to deal with the
inherent discreteness of galaxies (e.g., Szapudi \& Szalay 1997).
In what follows, the continuous version of the theory will be used 
for simplicity, and factorial moments are implicitly assumed wherever 
continuous moments are used in the spirit of the above.

Next, the mean correlation function, $\xiav = \avg{\delta^2}$,
and the amplitudes of the $N$-point correlation
functions, $S_N = \avg{\delta^N}_c/\avg{\delta^2}^{N-1}$, are computed
\footnote{$\avg{}_c$ refers to ensemble 
averages over the connected component}.
The galaxy correlation function measured via counts in cells is
actually a smoothed version of this function referred to as $\xiav$.
If the galaxy correlation function is a power-law then the average 
correlation function, $\xiav$, is a power-law as well with the same slope 
but with a different (increased\footnote{The amplitude is higher because
$\xiav$ is the average value of $\xi$ UP TO a given scale.}) amplitude 
that is determined with Monte-Carlo integration. 
For square cells and galaxy correlations with the usual slopes
($\gamma \sim 1.7$), the ratio $\xiav / \xi$
is typically less than a factor of 2. The calculation of
$\xiav$ is an alternative way to measure the two-point correlation
function and it is used in the computation 
of the normalization of the higher order
cumulants, i.e the $S_N$'s.

In the sequence of calculating the $S_N$'s from 
counts in cells via factorial moments, the most delicate
step is the actual estimation of the distribution of counts
in cells in the survey.
We begin by transforming celestial coordinates into equal area Cartesian 
coordinates to assure proper handling of the curvature of the survey 
boundaries on the sky. 
We then adopt a series of masks (\eg around bright stars)
to denote regions in the survey that should be excluded from analysis. These 
masks were defined during the object detection phase and help minimize the 
inclusion of spurious detections.
The infinitely oversampling method of (Szapudi 1997)
was used to estimate $P_N$ in square cells.
This method completely eliminates the measurement error 
due to the finite number of sampling cells. It can be 
sensitive, however, to edge effects on larger
scales. When all the survey masks are used,
only small scales up to $0.16^\circ$ could be studied because
larger cells would nearly always contain an excluded region. As a sensible
alternative, we eliminated all but 
the top 5\% of the largest masks and repeated
the analysis. This procedure extended the dynamic range of the 
angular scales probed and typically does not significantly alter the
results on smaller scales. In addition to shrinking the possible
dynamic range available to our measurement, the large number
of masks compromise the accuracy of the error estimate as well:
the geometry of the survey (as explained in more detail later)
could not be taken into account at the level of including the
geometry, position, and distribution  of the masks.

\subsection{Three-point correlation function}
\label{sec:3pcf}

The moments of counts in cells (i.e., the $S_N$'s) are the simplest descriptors
of the non-Gaussianity of a spatial distribution, both in terms
of measurement and interpretation. However, there are alternative statistical
measures, such as the $N$-th order correlation functions, 
that reveal more information about the galaxy distribution
by incorporating information about the geometry of higher order correlations. 
Indeed, the $S_N$'s are just higher order
correlation functions that have been integrated over a portion of 
the available configuration space, i.e. over the $N$-points within a given
cell.  Although averaging over the configuration space 
suppresses noise, it also erases some important information that
is, by contrast, kept by the $N$-th order correlation functions.
Specifically, the validity of the hierarchical assumption can be 
tested directly
(e.g., the first term in equation \ref{eq:expansion})
and ultimately the effects of bias and gravitational instability can be
distinguished. It is therefore desirable to compute as many of 
the $N$-th order correlation functions as the survey volume will allow.
Three-point correlation functions have been calculated for the 
Zwicky, Lick, and Jagellonian galaxy samples by Peebles \& Groth (1975), 
Groth \& Peebles (1977), and Peebles (1975), 
for Abell clusters by T\'oth, Holl\'osy, \& Szalay (1989, hereafter THS), 
for the APM galaxy survey by Frieman \& Gazta\~naga (1999), 
and for an IRAS-selected galaxy sample by Scoccimarro \etal (2000; in this 
case the calculation was performed in Fourier space). Only the
measurements by Peebles and Groth were performed on scales comparable
to us, the rest of the work concentrated on much larger angular scales.

A larger survey is needed to directly measure the three--point correlation 
function, $\xi_3 = \zeta$, over a large dynamic range in scale
than that required for the measurement of skewness, $S_3$. Physically,
this can be understood: the total number of triplets available in the 
survey are divided into finer subsets in the case
of $\xi_3$ than for $S_3$, in order
to examine a larger number of parameters. A smaller number of
triangles available for each bin of the three-point correlation
function naturally causes larger
fluctuations and, hence, larger statistical errors.

Szapudi \& Szalay (1998, hereafter SS) proposed a set of new estimators
for the $N$-point correlation functions using either
Monte Carlo or grid methods. We denote these
series of estimators SS$_{RN}$ and SS$_{WN}$, respectively.
They have as their primary advantage the most efficient
edge correction of any estimator developed to date.
The two-point correlation function was estimated both 
with the Landy-Szalay estimator (Landy \& Szalay 1993; hereafter LS), 
$(DD-2DR+RR)/RR$, and the grid indicator SS$_{W2}$. We employ
both methods primarily to verify that the SS$_{W2}$ results
agree with those derived from the well-tested and widely used LS estimator.
The three-point correlation function was
estimated by the grid indicator SS$_{W3}$.

The estimators SS$_{WN}$ ($N=2,3$) were implemented as follows.
If $D$ denotes the data counts in a sufficiently
fine grid, and $W$ is the characteristic function
of the grid, i.e. taking values of  $1$ inside the valid survey
boundary, and $0$ elsewhere, the  SS$_{W2}$ estimator for
the two-point correlation function is 
\begin{equation}
  \omega_2 = (DD-2DW+WW)/WW,
\end{equation}
where $DD$, $DW$, and $WW$ denote the data-data, data-window,
and window-window correlation functions. This is analogous
to the LS estimator.
The SS$_{W3}$ estimator for the three-point correlation function is
\begin{equation}
  \omega_3 = (DDD-3DDW+3DWW-WWW)/WWW.
\end{equation}

The key difference between the LS and SS$_{W2}$ estimators is that
the latter statistic is analogous to Euler's method instead
of Monte Carlo integration. 
In fact, SS$_{RN}$ is the direct
generalization of the relation for the two-point function, which is
LS $=$ SS$_{R2}$. For the grid estimators
SS$_{WN}$  the random catalog R is replaced with the characteristic 
function of the survey estimated on a grid, $W$. The difference amounts
to a slight perturbation of the radial bins.

The main advantage of using a grid instead of a random catalog is 
computational speed, which starts to become prohibitive for the
three-point function (at least with current technology computers
and the present algorithms).
The CPU time for calculating two-point correlation functions
with the LS estimator is on the order of hours for about $n = 10^5$
galaxies, and it roughly scales as $n^2$. For the three-point
correlation function $n$-times more CPU power is thus
needed, \ie it would take $\sim10^5$ hours on a fast
workstation to calculate the SS$_{R3}$ estimator. The SS$_{WN}$ series of
grid estimators scale again as $n^N$, but here $n$ is the number
of grid points. A significant speed up was achieved for the
SS$_{W3}$ estimator by storing precalculated distances of the 
grid-point pairs in a look up table; population of the look up table
required $\simeq n^2$ operations. This did not change the
scaling of the actual three-point estimator, but did minimize CPU
time as floating point operations are eliminated from the main
calculation. With this method,
the three-point function for all magnitude cuts could be
calculated in about a CPU day on a $256^2$ grid. 

For the two-point correlation function, where it was
computationally feasible, we confirmed that SS$_{W2}$ produced 
results that were consistent with the LS estimator on 
scales larger than a few grid spacings, as it should.
To repeat the same test for the three-point function 
would have necessitated supercomputer resources (see discussion above).
This was not deemed justifiable as it is clear that no
qualitative differences are expected for the three-point function
from the slight perturbation of the binning introduced by
the grid estimators.

We computed $\omega_3$ for all magnitude slices up
to about 1 degree scales. The star-galaxy classification accuracy
degrades at faint magnitudes ($I \gsim 22$) in ways that are detailed
in paper I. In brief, a fraction of the faintest
compact galaxies are often classified as stars. The statistical
correction for this effect is to include some faint ``stars" into the
galaxy catalog based upon a selection function that is derived from an
extrapolation of the star counts at brighter ($I < 20.5$) magnitudes where
classification is extremely reliable. The effect of the correction is to
restore, at a statistical level, the correct galaxy-to-star ratios at
faint magnitudes. While this introduces some minor contamination
of the galaxy sample by stars, the stars are presumably randomly distributed
on the angular scales being considered. Therefore, except
for the highly improbable case when a star cluster is included, any stellar 
contamination dilutes the correlations by the usual factor $f_s^2$, 
where $f_s$ is the fraction of stars in the survey. At the magnitudes 
where the statistical classification corrections are required ($I > 21.5$), 
the number of stars introduced into the galaxy catalog as a result of the 
correction is estimated to be less than 10\% of the galaxy population. 
The effect of the dilution is, thus, negligible.

An alternative approach to dealing with faint object misclassification
is to only include in the analysis those objects initially
classified as galaxies. If the misclassified galaxies have the 
same clustering properties as the correctly classified galaxies then the
correlation functions will not change significantly
(small changes may occur as a result of cosmic variance).
However, if the compact objects likely to be misclassified as stars
cluster differently, their omission could introduce a 
bias. In that case, sampling the objects classified as stars
statistically according to an estimated misclassification ratio, as discussed
above, is the remedy.

Lastly, one can compute the correlation functions using all detected
objects {\em regardless} of their classification. In this case,
one can again estimate the ``true" galaxy correlation function by
extrapolating the bright stellar number counts and multiplying the
correlation amplitude by the factor $N_{Obj}^2/(N_{Obj}-N_{Star})^2$
where $N_{Obj}$ is the total number of objects in a given magnitude
bin and $N_{Star}$ is an estimate of the number of stars in the same bin. 

To determine if misclassification effects are a significant source of bias,
we estimated the two-point correlation function using all three strategies 
in all slices. The four brightest slices ($17 \le I < 21$)
exhibit no discrepancies between any of the methods, indicating that
misclassification is negligible in this flux range.
For the deepest slice ($22 \le I < 22.5$), 
the results based on the statistical inclusion of
faint stars according to an estimate of the misclassification rate is in 
excellent agreement with the corrected correlation function derived using
all detected objects. Agreement with the results using only objects 
classified as galaxies was marginal. 
The $21 \le I < 22$ slice is an intermediate case. 
These results suggests that correction for
misclassification is essential in this survey for $I \gsim 21.5$. 

For the deepest slices, we have thus performed the estimation 
of the three-point correlation functions using the first statistical 
correction discussed above (results from this method are denoted
with the letter ``s" in Tables~1 and 3)
as well as an estimation based on using only
those objects classified initially as galaxies.
The difference in the results between the two methods provides
an upper estimate of the systematic error introduced
by star-galaxy misclassification. For the two
deepest slices ($21 \le I < 22,\ 22 \le I < 22.5$), 
the statistical procedure should yield the correct results.

\section{Results}
\label{sec:results}

\subsection{Counts in Cells}
\label{sec:cic}

\begin{figure}
\centerline{\hbox{\psfig{figure=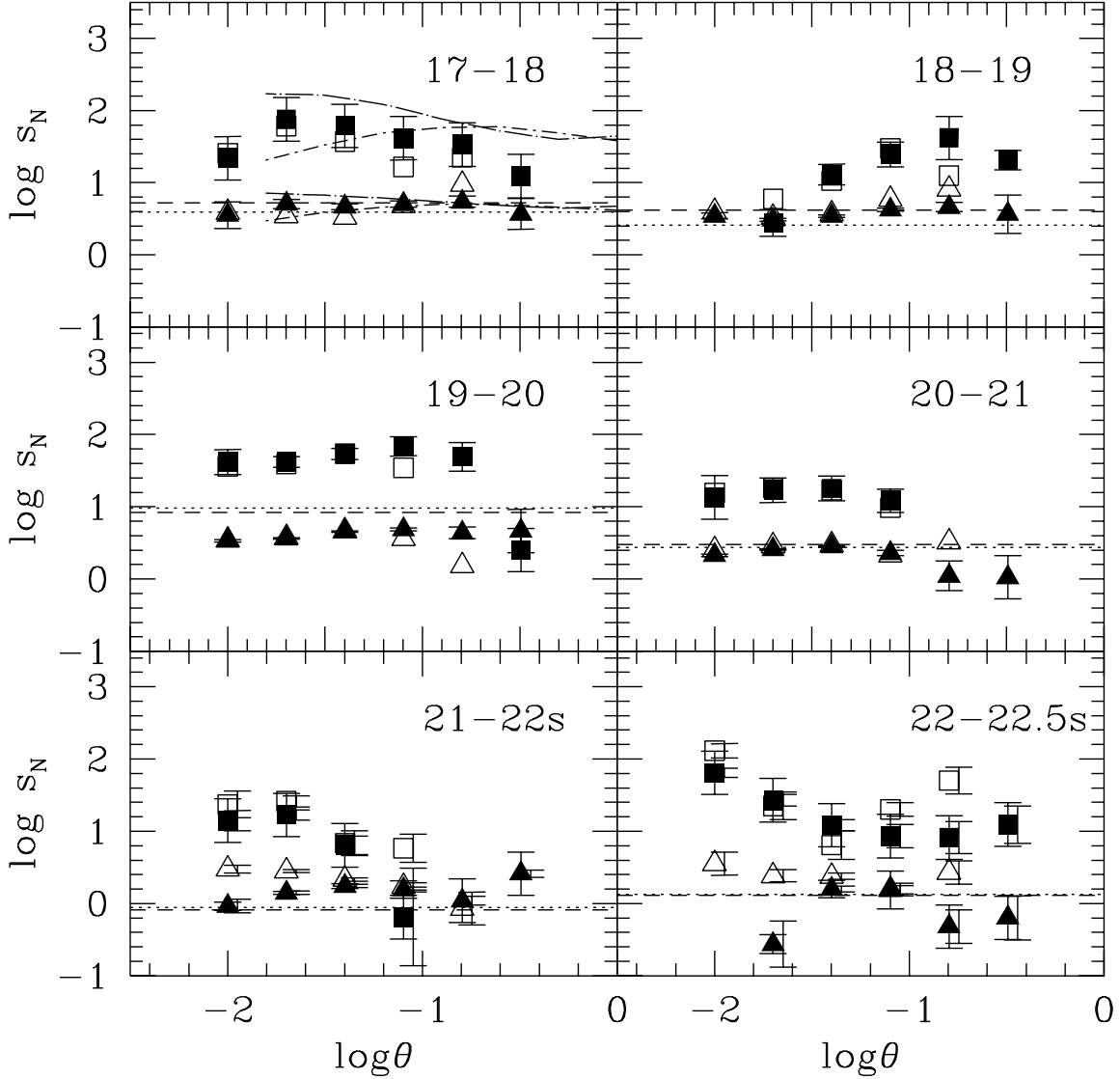,width=16cm}}}
\caption{$s_3$ and $s_4$ estimated from the Deeprange
are shown with triangular and rectangular symbols,
respectively. Open symbols are the results when 
all survey masks are used.
Closed symbols display the results when only the
top 5\% largest masks are used. Dotted
and dash lines correspond to a $\chi^2$ fit of the
three-point correlation function with third order
terms free or locked to zero, respectively.
The error bars are estimated using the non-linear
FORCE package. Systematic error estimates are also shown for the
faintest ($I > 21$) results as right-shifted error bars. 
The dash-dot and long dash-dot curves display
the results of the APM and EDSGC surveys, respectively, in the
upper left ($17 \le I < 18$) window.}
\label{fig:s3s4}
\end{figure}

The $s_3$ and $s_4$ results are shown in Figure~\ref{fig:s3s4}.
In what follows lower case letters, 
such as $s_N$, denote the observed angular quantities and
uppercase characters, such as $S_N$, represent the deprojected, 
spatial statistics.
In each plot, open symbols show the exact estimates
when all cells containing survey masks are excluded (hereafter referred to as
the measurement type ``E") and closed symbols show the values
when only cells containing the largest 5\% of the masks were excluded
(hereafter referred to as measurement type ``M").
The magnitude cuts are shown in the lower right hand corner of each
plot.  The triangular symbols
refer to $s_3$ while the rectangular ones to $s_4$.
The $s_N$ data ($N = 3,\ 4$) and their
estimated uncertainties are given in Table~1. 
The results in Table~1 for $I < 21$ are for the ``M" type measurements whereas
those for $I \geq 21$ are for the ``E" type measurements.
We provide the ``E" type measurements for the faintest slices because they
provide the best fidelity and accuracy in the presence of any faint end
systematic errors in the catalog at the expense of angular scale dynamic range
(see discussion below).

For the scales considered, deprojection was performed using
the hierarchical Limber's equation:
\begin{equation}
s_N = R_N S_N.
\end{equation}
(see e.g., Szapudi, Meiksin, \& Nichol 1996).
Table~2 lists the $R_N$'s for a typical
choice of the luminosity function. The $I-$band
luminosity function is assumed to be the
same evolving Schechter function as in paper I with
$M^*(z) = -20.9 + 5 log h -\beta z$, $\alpha = -1.1$. Our fiducial
parameters for the deprojection are $\beta = 1.5$, and
$\gamma = 1.75$ for the slope of the two-point spatial correlation function.
The cosmological parameters adopted are $h = 0.65$,
$\Omega_o = 0.3$, and $\Lambda = 0$.
The results are independent of the normalization of the luminosity function,
and robust against variations in these parameters within the
accepted range, or even beyond.
For instance, the effect of varying all parameters
simultaneously over the ranges
$\beta = 1-2,\ \gamma = 1.5-1.75,\ h = 0.5-0.75,\ \Omega = 0.3-1$
is a $\sim10-15$\% change in $R_3$.
A non-zero cosmological constant $\Lambda$ was not considered; the effect
of a reasonable non-zero $\Lambda$ on our estimates of the higher order
correlation statistics is expected to be bracketed by the changes in $\Omega$
tested above. Table~2 also gives $R_3$ for $\beta = 1$
and for $\Omega = 1$ to explicitly
demonstrate the relatively weak dependence of the
projection coefficients on these parameters.

The problem with the ``E" type measurement is that the large number of 
excluded cells severely limits the maximum scale for which we can
derive constraints.  Indeed, on scales above 10 arcminutes any 
cell would intersect a masks with almost unit probability. 
With ``M" type measurements, we can increase the maximum spatial scale
by a factor of 2 at the price of contaminating the estimate slightly.
The typical effect of such a contamination is a slight decrease in
the $s_N$'s, although it depends on the exact spatial and size
distribution of the masks and, as such, is extremely complicated to 
correct for or predict. The complex distribution and approximately
power law size distribution of the masks prevented the application
of counts in cells on larger scales, still relatively small compared
to the characteristic size of the survey. This unfortunate phenomenon
can be understood in terms of the masks effectively chopping up the
survey into a large number of smaller surveys. Nevertheless, the
large area of the Deeprange survey is still essential for the calculation
of higher order statistics, as otherwise the finite volume errors
would render the measurement of the $s_N$'s meaningless, even on
the present small scales.
However, the contamination decreases as the scale increases, since
the area-ratio of the ignored mask is decreasing,
and, thus, if the ``M" and ``E" type measurements agree out to a given
angular scale, then the ``M" measurements on larger scales
can be considered relatively free of contamination whereas the
``E" measurements may show edge effects. The two faintest slices, however,
exhibit some disagreement between the ``E" and ``M" measurements on the 
smallest scales (see Figure~\ref{fig:s3s4}) and therefore it is 
the ``E" measurements that should be considered
the most reliable results in these cases.

\subsubsection{Error Estimation}
\label{sec:errors}

The errors of the measurements (columns 4, 6, 9, and 11 in Table~1)
are estimated using the full non-linear error calculation
by (Szapudi \& Colombi 1996, Szapudi, Colombi, \& Bernardeau 1999), 
obtained from the FORCE (FORtran for Cosmic Errors) package. 
All the parameters needed for
such a calculation were estimated from the survey
self-consistently. These are the perimeter and area of the survey,
the cell area, the average galaxy
count and the average two-point correlation function
over a cell, and the average two-point correlation 
function over the survey (estimated from the integral
constraint fits of paper I). The
higher order $S_N$'s up to 8 were needed as well. They
were calculated from extended perturbation theory
(Colombi \etal 1997; Szapudi, Meiksin, \& Nichol 1996). 
This theory is defined by one parameter,
$n_{eff}$, which is identical to $n$, the local slope of the 
power spectrum on weakly non-linear scales, and is purely 
phenomenological on small scales. It can be determined from $S_3$
according to Equation~\ref{eq:sqbias}, and was 
measured from the average $S_3$ estimated
from the survey itself. In addition, the package uses a model for
the cumulant correlators, the connected joint moments
of counts in cells. The model by Szapudi \& Szalay 1993
was chosen, as this was shown to provide a good description
of projected data (Szapudi \& Szalay 1997).

To estimate the uncertainties in the $S_N$'s, the FORCE package
uses a perturbative procedure that is numerically accurate for small errors.
When the fractional errors become  $> 1$, the value computed
is meaningless and only indicates, qualitatively, that the errors are large.
In those instances we renormalized the plotted error to $100\%$ 
and the corresponding entry in Table~1 is given as ``L''.

The accuracy of the fully non-linear error calculation method
has been thoroughly checked (see Szapudi, Colombi, \& Bernardeau 1999;
Colombi \etal 1999). However, 
one cautionary note is in order: although the calculation 
takes into account edge effects to second order (the ratio of the survey
area to survey perimeter), the use of masks introduces complex
geometrical constraints and extra edge effects.
This is the main reason why the counts in cells method is limited to fairly
small scales; the errors associated with these effects are not 
accurately modeled by the FORCE.

For our two faintest bins ($I > 21$), we also provide an estimate of the
fractional systematic errors associated with uncertainties in the
statistical correction for object misclassification. These are given as
the second error value in the parentheses in Table~1. The systematic errors
are just the standard deviations in the mean $s_3$ and $s_4$
computed from 10 realizations of the statistical misclassification
correction described in \S\ref{sec:3pcf}. 
The fractional systematic errors are shown by the right shifted
error bars in Figure~\ref{fig:s3s4}. Systematic errors 
for the brighter bins are negligible ($\lsim 2$\%). 

\subsection{Three-point correlations}

\begin{figure}
\centerline{\hbox{\psfig{figure=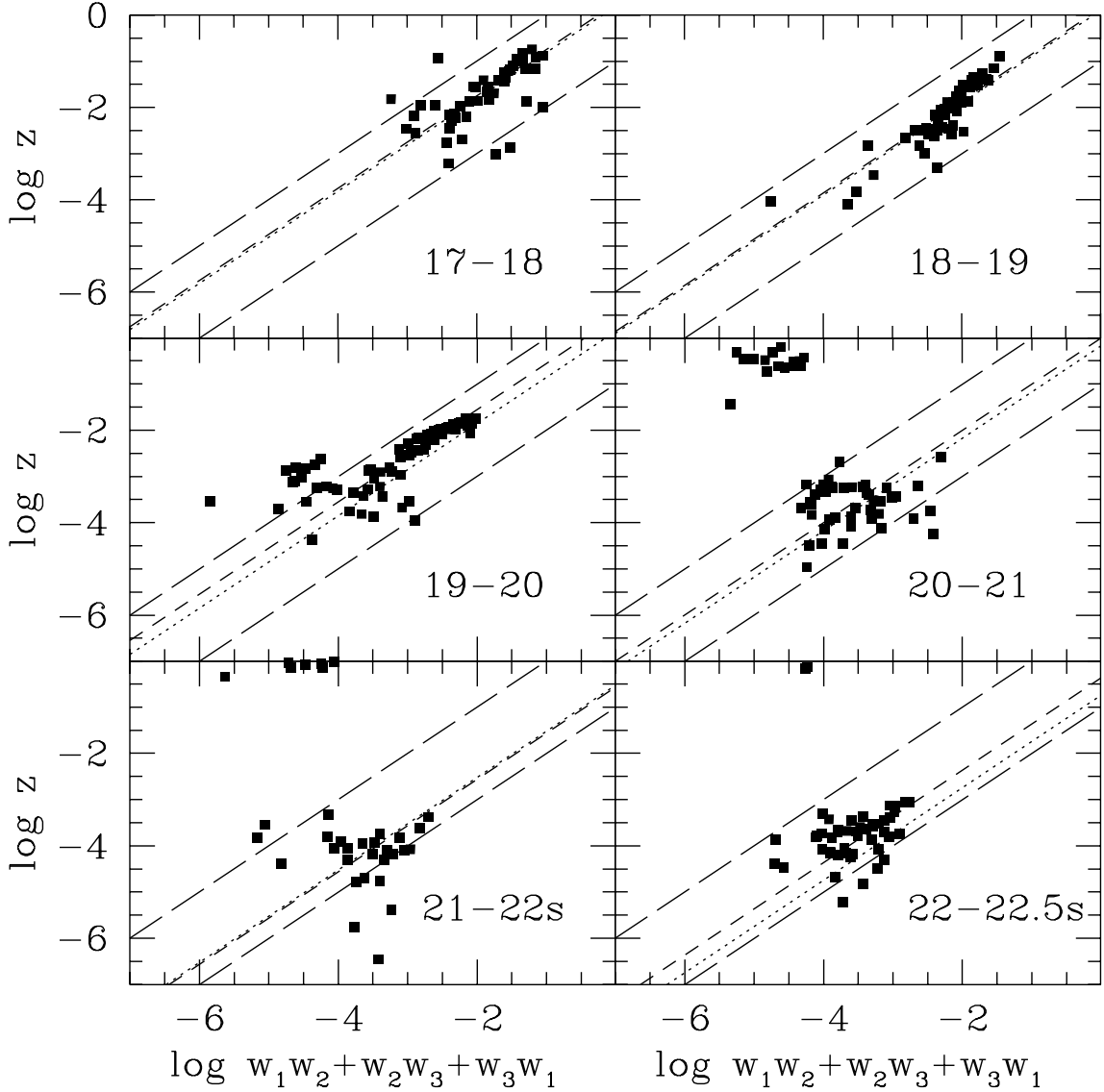,width=16cm}}}
\caption{The three-point correlation function $z$
is displayed in terms of the hierarchical term
$w(1)w(2)+w(2)w(3)+w(3)w(1)$.
The long dashes show the limits that were used to
exclude outliers in the $\chi^2$ fit for the parameters
of the general third order connected expansion. Dotted
line displays the results from the counts in cells analysis,
while short dashes show the $q_3$ from the $\chi^2$ fit.}
\label{fig:w3}
\end{figure}

Figure~\ref{fig:w3} shows our measurements
of the angular three-point correlation function, $z(a,b,c)$, 
as a function of the hierarchical term 
$\omega(a)\omega(b)+\omega(b)\omega(c)+\omega(c)\omega(a)$,
where $\omega$ is the two-point correlation function. The magnitude ranges
are shown in the lower right hand corner of each subplot.
The three-point correlation function was estimated in logarithmic
bins for each side of the corresponding triangle. The figure
makes no attempt to display shape information, although it is available
from our estimator. 
The interpretation of projected shape space is extremely difficult 
because several different three-dimensional
triangles project onto identical angular configurations. 
In the weakly non-linear regime,
it is possible to project down the firm predictions
for triangular configurations from a given theory
and compare them with the observed angular three-point correlation function
(e.g., Frieman \& Gazta\~naga 1999). For the small scales we are
considering, however, a typical shape would deproject to a mixture
of triangular galaxy configurations that are in the highly and 
mildly non-linear regimes. Therefore, we have chosen a
different, entirely phenomenological, approach pioneered by
THS. This essentially
consists of fitting the three-point correlation function with
a class of functions motivated by a general expansion; 
details of the fitting procedure and error estimation are 
presented in Appendix A.

Table~3 summarizes the results of the three-point correlation 
function fitting procedure. Column 2 of this table gives the
number of degrees freedom in the fit. We perform fits
with the third order terms ($q^{111}$ and $q^{21}$) 
free or locked to zero - hence the two
different sets of results for each magnitude bin.
On scales where the correlations are small ($\xi_2 \lesssim 10^{-4}$)
and the relative fluctuations of the estimator are becoming
increasingly large, the division by the hierarchical term (see Appendix A)
becomes unstable and produces a multitude of outliers
that are easily identified in Figure~\ref{fig:w3}. From
the measurement of $S_3$ from Deeprange, $q_3$
can be safely bracketed with $0.1 < q_3 < 10$, at least for
the brighter magnitude cuts.
These are fairly conservative limits and are in agreement with
other measurements (e.g., Szapudi, Meiksin, \& Nichol 1996) as well.
These limits are displayed as long dashes
on the figures representing three-point correlation functions,
and were used in the fits to eliminate outliers.
For a few of the deepest slices the fit could be sensitive 
to the exact placement of the lower $q_3$ outlier cut. 
All the fits were, thus, reevaluated with the limits $0.01 < q_3 < 10$
as well. These extended fits are given in columns 7 through 10
in Table~3. The $\chi^2$ values in columns 3 and 7 of 
Table~3 are the full (unreduced) values. 

Any significant difference between the fits in Table~3 for
a given magnitude interval suggests large systematic errors.
All the fits were performed by standard computer algebra
packages, and the parameters were found to be robust
with respect to the initial value assigned to them.
The results in Table~3 can be compared with those in Table~1 
using the fact that $s_3 \simeq 3 q_3$.
This equation is not exact, because the shape of the cell
influences the integral performed in the definition of $s_3$.
The resulting form factors amount to only $\simeq 2-3\%$, which
is the accuracy of the approximation (Boschan, Szapudi \& Szalay 1994).
Our fitting procedure cannot yield an accurate error on the
fitted parameters, since the input errors for the $\chi^2$ were
not accurately determined. Nevertheless, the FORCE error bars
should give a good indications from the previous tables for
the expected statistical uncertainty since the hierarchy is
approximately true. 
Furthermore, any differences in the results introduced either by
neglecting the statistical correction for star/galaxy misclassification
or by altering the $q_3$ fit limits (see Table~3) 
should give a reliable indication of the systematic errors.

We did not attempt to compute systematic errors for the
three-point correlation function because it would have been very
time consuming computationally. We note, however, that one can gauge
the amplitude of any systematic errors by comparing the results
with and without the statistical misclassification correction
in Table~3. Furthermore, the realization of the
misclassification-corrected galaxy catalog used in the
three-point correlation function computation has 
$s_3$ values that are very similar to the values derived
from the counts-in-cells method.

\section{Discussion and Summary}
\label{sec:sum}

Our constraints on the higher order clustering of the
brightest galaxies in our survey are in good agreement with previous work. This
can be seen in Figure~\ref{fig:s3s4} for our $17 \le I < 18$ subsample
where we also display the results reported by
Szapudi \& Gazta\~naga (1998, see also Gazta\~naga 1994) 
for the APM $17 \le b_j < 20$ galaxy sample and those reported
by Szapudi, Meiksin \& Nichol (1996) for the EDSGC with 
dash-dots, and long dash-dots, respectively. 
Comparison of our $17 \le I < 18$ subsample with the above larger surveys is
justifiable because the projection coefficients only vary slowly with depth.
The APM and EDSGC measurements are the present state-of-the-art 
for shallow angular surveys, to be
superseded only with the Sloan Digital Sky Survey. The APM and EDSGC
measurements agree with each other on intermediate-large scales
(see Szapudi \& Gazta\~naga 1998 for detailed comparison
of counts in cells measurements of the two surveys). On small scales,
however, the EDSGC results are higher than that of the APM.
The discrepancy between the APM and
EDSGC results can most likely be attributed to differences between the 
deblending procedures used in the construction of the two catalogs
(Szapudi \& Gazta\~naga 1998). Our measurements, 
especially for $S_3$, appear to lie mostly between the two where they disagree
and consistent with them, although perhaps
somewhat lower, where they agree. We conclude that the
Deeprange counts-in-cells measurements are in agreement
with previous estimates in shallow angular surveys, despite
the difference in wavelength and the smaller area.
Constraints from the three-point correlation function mirror the above results
as well -- our measurement of $Q_3 = 1.57$ from the
$17 \le I < 18$ subsample is remarkably close to that of Groth
and Peebles (1977), $Q_3 = 1.29 \pm 0.21$ compiled
from the Zwicky, Lick, and Jagellonian samples. 

For the collection of magnitude limited cuts,
$S_3$ and $S_4$ are approximately scaling. The $S_N$'s
appear to decrease with depth, which suggest a small
evolution with $z$, even though the redshift
distribution of each slice is fairly broad.
This is in accordance with theories of structure formation
where the initial Gaussian fluctuations grow under the influence
of gravity (for perturbation theory see e.g.,
Peebles 1980; Juszkiewicz, Bouchet, \& Colombi 1993;
Bernardeau 1992; Bernardeau 1994);
for $N$-body simulations see e.g., 
Colombi \etal 1995; Baugh, Gazta\~naga, \& Efstathiou 1995;
Szapudi \etal 2000b),
and where there is a small bias.

\begin{figure}
\centerline{\hbox{\psfig{figure=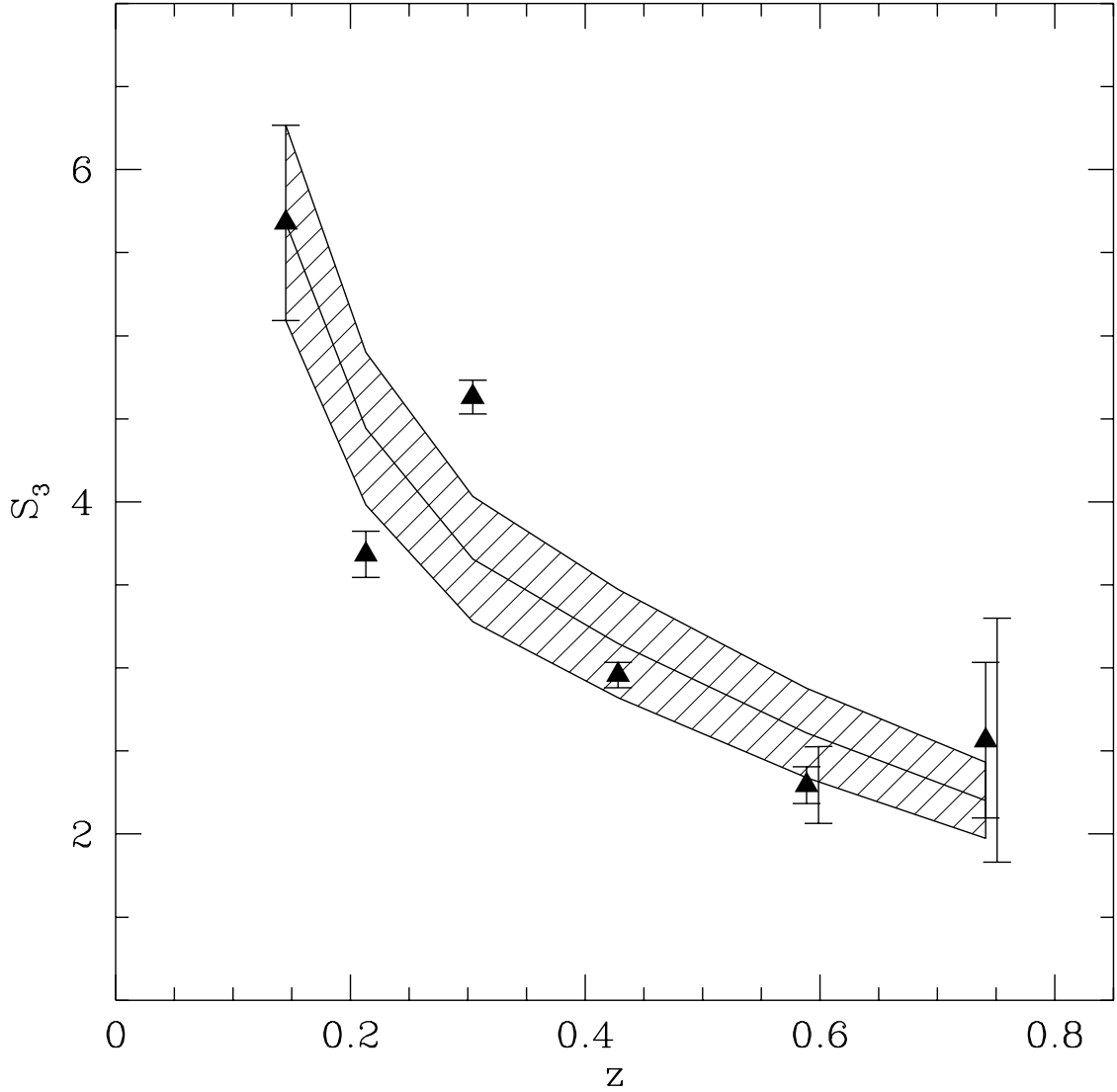,width=16cm}}}
\caption{Constraints on the time evolution of biasing are shown. 
The open symbols are the predicted
$s_3$ from a simple model (see \S\ref{sec:sum} for details).
The solid symbols display $S_3$ measured at $0.04^\circ$, where the
FORCE error bars are the smallest. Strictly, 
each $S_3$ is at a slightly different scale in comoving
$\mpc$, but because of the flat scaling of $S_3$ this is not
an important effect. The centered error bars on all data points
are those derived from the FORCE package. The right shifted error
bars for the two faintest data points are the FORCE$+$systematic
errors (see \S\ref{sec:errors} for details).}
\label{fig:bias}
\end{figure}

The wide area of the Deeprange survey enables a self-contained
study of the time evolution of the bias -- there are a statistically sufficient
number of low-$z$ galaxies contained in the survey that
a self-consistent constraint can be derived. Our measurement of the
the bias evolution uses a simple model based on the 
parameterization of the evolution
of the two-point correlation function discussed in
paper I. Briefly, we parameterize the redshift dependence 
of the correlation function as
\begin{equation}
\xi(r,z) = (\frac{r}{r_0})^{-\gamma} (1+z)^{-3-\epsilon},
\end{equation}
where $\gamma$ and $r_0$ are the slope and correlation length,
and $\epsilon$ is a phenomenological parameter describing
the evolution (e.g. Peebles 1980; Efstathiou \etal 1991;
Woods \& Fahlman 1997). 
For $\Omega = 1$ and linear evolution, the correlation function 
at any $z$ can then be compared to that at the current epoch
through the mapping $(1+z)^2\xi(r/(1+z),z)$. Although this mapping
is strictly for $\Omega = 1$, variations
in the mapping due to scenarios with $\Omega < 1$ are well within
the range of uncertainties due to the exclusion of all details
associated with specific galaxy formation processes in this simple
model. Under the {\em same} mapping the
$S_N$'s would be invariant. Therefore, we can define the ratio
of the mapped correlation function to that at the current epoch as 
\begin{equation}
  b(z)^2 = (1+z)^{\gamma-1-\epsilon}.
\end{equation}
Since our higher order measurements probe galaxies in the highly non-linear
regime, it is appropriate to use Equation~\ref{eq:sqbias} for biasing,
which yields $S_3(z) = S_3(0)/b(z)^2$. Figure~\ref{fig:bias} displays
the predicted $s_3$, normalized using the measurement from the $17 \le I < 18$
sample, and the actual measurements at $\theta = 0.04^\circ$
(solid symbols). Redshifts are assigned to the observed data by computing
the median $z$ for each magnitude slice as predicted by 
the $\beta = 1.5$ LF evolution model
described in \S\ref{sec:cic}. The predicted data were computed using
$\gamma = 1.75$, and $\epsilon = -1$.
Strictly, $S_3$ should be compared
at the same comoving scale, but the flat scaling
of the $s_3$ allows comparison at the same angular
scale instead. This model is clearly an oversimplification
of the time evolution of the bias, and yet, it captures the trend
presented by the data remarkably well. In addition, the time
evolution of the bias is constrained, according to
Equation~\ref{eq:sqbias}, to be less than a factor of 2 between the current
epoch and $z \sim 0.75$.

On the other hand, the above observations are
in strong contrast with expectations from non-Gaussian scenarios. 
In these models, primordial non-Gaussianity (e.g., skewness) 
grows in linear theory. For Gaussian initial
conditions, the growth of the skewness is a second
order effect. Thus according to
theoretical calculations (Fry \& Scherrer 1994), and simulations
(Colombi 1992), the $S_N$'s should have been larger in the past
in non-Gaussian models compared to their 
Gaussian counterparts.  As an example, $S_3$ is expected to be 
a factor of 2 larger at $z \simeq 1$ than at $z \approx 0$
for the typical initial conditions in cosmic string
models (Stebbins 1996, private communication; Colombi 1992). 
The growth of $S_4$ with increasing $z$ is
expected to be even more prominent. These effects would 
have been detectable in our survey, despite the dilution effects
of projection and possible systematic errors at the survey magnitude limit. 
Formally, the non-Gaussian expectation is about $7-8\ \sigma$ from 
our measurement at the highest redshift. Even generously
doubling our error bars (which would make the previous 
naive biasing model based on Gaussian initial conditions a
perfectly reasonable fit to the data in Figure~\ref{fig:bias}) 
would still exclude typical cosmic string
non-Gaussian initial conditions at about a $3.5-4\ \sigma$ level.
The Deeprange data, thus, strongly favor Gaussian initial conditions.

While string initial conditions are not favored for many
reasons (e.g., Pen, Seljak, \& Turok 1997; Albrecht, Battye, \& Robinson 1998), 
our expectation is that the above arguments
would hold for a large class of non-Gaussian models. However, it is possible
to invoke biasing schemes that are in accord with our observations but
that also mask the signature of non-Gaussian initial conditions. 
For example, strong bias at early
times decreases the higher order moments, counteracting the 
effects of the initial non-Gaussianity. Later the bias
naturally decreases, resulting in an increase of the $S_N$'s. 
While perhaps such a scenario is not completely unimaginable physically,
e.g. by invoking a strong feedback during galaxy formation,
it would require unnatural fine tuning
in order to assure that the time evolution of biasing effectively cancels
the time evolution of the non-Gaussian initial conditions.
Thus, while our results cannot rule out initial
non-Gaussianity with high certainty, the most natural explanation is
that Gaussian initial conditions of the fluctuation field
grew via gravity. Galaxies appear to trace mass
quite accurately, and the small evolution of bias predicted
by our naive model appears to describe the trends of the
data fairly well.

The three-point correlation function appears to be hierarchical down to
$I = 22.5$. Our estimates of the non-hierarchical term, $Q^{21}$, are uniformly
small in amplitude (see Table~3),  
although a cubic term $Q^{111}$ cannot be excluded
with high significance. While the $\chi^2$ analysis is 
only approximate in nature (the simple error model
did not attempt to estimate bin-to-bin cross-correlations),
the inclusion of a cubic term does not typically result
in a significant change in the goodness of fit.
Gravitational instability predicts $Q^{111} \simeq 0$,
thus a small cubic term is likely to mean mild bias,
as predicted by general bias theory. In the case
of a Gaussian field with a completely general
bias function $Q^{111} = Q_3^3$ is expected under fairly
general conditions (Szalay 1988). Since we find $Q^{111} \ll 1$
and $Q_3 \gtrsim 1$, the galaxy density field cannot be
a biased version of a Gaussian field but, rather, a mildly
biased version of an underlying non-Gaussian field.
We emphasize that the non-Gaussianity being referred to here 
is that which is induced by non-linear gravitational amplification 
and does not refer to the nature of the initial perturbation spectrum.
This interpretation is subject to cosmic variance
on $Q^{111}$, possible systematic errors, and possible
stochasticity of the bias, all of which could contribute
to the cubic term.

The best fitting $S_3 = 3\times Q_3 \simeq Q^{11}$ from the three-point
correlation measurements are also displayed in Figure~\ref{fig:s3s4}
as two horizontal lines: the
dotted lines show result of the full fit, while dashed lines
display the results of the restricted fit. 
Despite the fact that the three-point correlation formula extends
to larger scales than the counts in cells analysis, the best fitting 
$Q^{11}$ appears to be in excellent agreement with $S_3$ obtained from 
counts in cells, with the exception of the $19 \le I < 20$ results where 
there is a factor of two discrepancy. The generally good agreement between
the two methods, however, is a further indication of 
the insignificance of the cubic term $Q^{111}$.
In Figure~\ref{fig:w3}, the dashed lines show
the best fitting $Q^{11}$ and the dotted lines display $Q_3 = S_3/3$ as 
obtained from the counts in cells analysis. This demonstrates 
the agreement between the counts in cells analysis and the direct
three-point correlation function estimation from a 
different perspective. The disagreement seen above for the $19 \le I < 20$
sample is less significant here and, thus, 
it must stem from the smallest scales. The points display
a slight non-hierarchical curvature as well. More accurate
measurements from even larger surveys will be required to assess whether
this is a significant behavior.

In summary we have measured moments of counts in cells
and the three-point correlation function in the Deeprange
survey. These constitute the deepest higher order clustering
measurements to date. The moments measured on small scales appear to be
hierarchical, and the three-point function, extending to
larger scales continues this hierarchy. While the cubic
term resulting from possible bias could not be excluded with 
high significance, the hierarchical assumption holds to a good approximation.
This argues that gravity is the dominant process in creating
galaxy correlations with bias having a detectable but
minor role. Qualitatively, models with Gaussian
initial conditions and a small amount of biasing, which increases slightly
with redshift, are favored. The large area of the Deeprange survey
allows us to study the evolution of bias over a relatively broad magnitude
range. We find that the bias between $I-$band selected galaxies 
and the underlying matter distribution increases slightly with increasing
redshift (up to $z \sim 0.8$) but not by more than a factor of 2.

\acknowledgments

In Durham, IS was supported by the PPARC rolling grant for 
Extragalactic Astronomy and Cosmology. The FORCE (FORtran
for Cosmic Errors) package can be obtained from its
authors, S. Colombi, and IS 
(http://www.cita.utoronto.ca/$^{\sim}$szapudi/istvan.html).
IS would like to thank Alex Szalay for stimulating discussions.

\appendix
\section{Fit for the Three-point Correlation Function}

The three-point correlation function can be expanded in terms
of powers of the two-point correlation function. This is called
a Meyer clustering expansion and it is commonly used in the field 
of statistical physics as well.  
The second (leading) order term in the expansion corresponds to the 
hierarchical assumption, which is predicted by gravitational
hierarchy (Peebles 1980). The higher order terms, which are
determined by the nature of the galaxy--matter biasing (Szalay 1988), 
represent corrections to the hierarchy that result in non-trivial shape
dependencies.

\begin{figure}
\centerline{\hbox{\psfig{figure=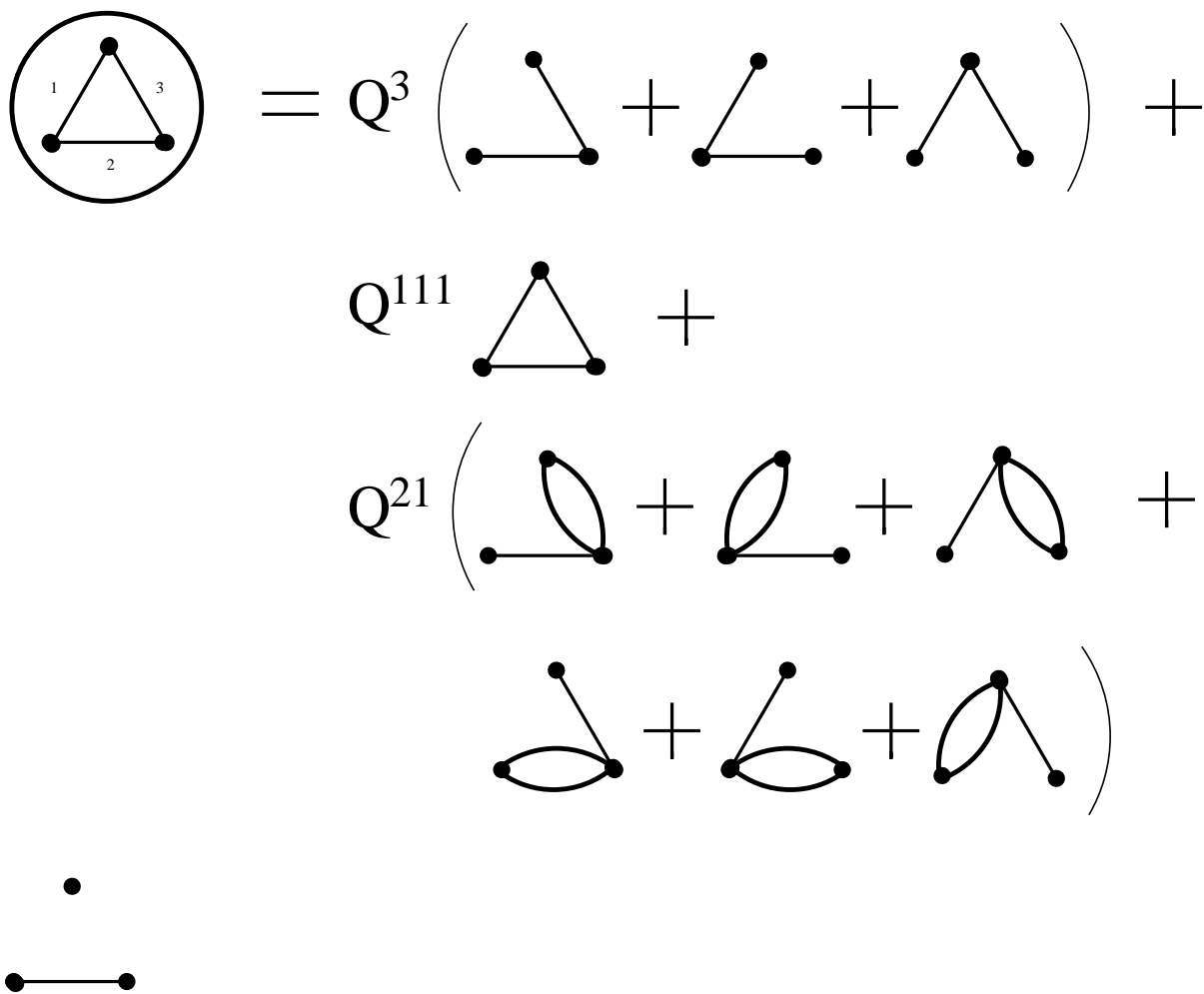,width=16cm}}}
\caption{Graphical representation of the Meyer cluster
expansion. The coordinates are denoted by $1,2,3$ in
the symbol for the three-point function, while
each vertex on the right hand side of the ``equation''
contain the possible tree, and one-loop connected
components of the expansion. The lower left corner
illustrates a disconnected term as an example, which is not
used in our expansion.}
\label{fig:Meyer}
\end{figure}

The Meyer clustering expansion, just like any spatial expansion,
can be represented by a Feynman-like graphical representation,
shown on Figure~\ref{fig:Meyer}.
This figure also illuminates the terminology of connected components,
since they correspond to connected components of the graph in
the pictorial representation.
By definition,  three-point function is
a connected third order moment: it is the extra probability
of a triangle above those predicted by Poisson and Gaussian
(i.e. two-point) terms. Consequently, it makes sense to use 
a connected Meyer expansion for it. 

Our aim is to project a full third order connected Meyer clustering
expansion of the three dimensional three-point function 
to its angular analog. Specifically,
the spatial three-point correlation function
can be expressed as a third order connected expansion
of the two-point correlation function, $\xi$, as
\begin{eqnarray}
  \xi_3 = \zeta(1,2,3) &=& Q^{11}\left(\xi(1)\xi(2)+\xi(2)\xi(3)+\xi(3)\xi(1)\right)+  \cr
               & &  Q^{111}\left(\xi(1)\xi(2)\xi(1)\right)+
                 Q^{21}\left(\xi(1)^2\xi(2)+sym \right).
   \label{eq:expansion}
\end{eqnarray}
This equation is directly translated to the graph on Figure~\ref{fig:Meyer}.
This expansion contains all the possible {\it connected} terms; THS included
all possible terms, including disconnected ones. 
Their notation is modified here slightly to avoid confusion with the 
cumulant correlator notation introduced since THS.

The projected three-point correlation function can be written as
\begin{equation}
  z(a,b,c) = \sum_{i = 11,111,21} q^i \omega_i(a,b,c),
  \label{eq:zmod}
\end{equation}
where $a,b,$ and $c$ are the angles of a triangle on the sky.
The $q^i$ terms are related to the three-dimensional $Q^i$ via 
$q^i = R_i Q^i$. For details see THS. 
The projection coefficient for the 
hierarchical term $R_{11} \equiv R_3$ is identical to that of
$S_3$. Since we have found that all other terms are
consistent with zero (i.e. the hierarchy is a good approximation)
we only deproject the hierarchical term.
The individual terms of equation \ref{eq:expansion} project down
to two dimensions as (THS)
\begin{eqnarray}
\omega_{11}(a,b,c) &&= \omega(a)\omega(b)+\omega(b)\omega(c)+\omega(c)\omega(a)\cr
\omega_{111}(a,b,c) &&= \left(\frac{\omega(a)\omega(b)\omega(c)}{a+b+c}\right)
\frac{\pi^2}{H(\gamma)^3}\left(\frac{180}{\pi}\right)\cr
\omega_{21}(a,b,c) &&= \left(\frac{\omega(a)^2\omega(b)+\omega(a)\omega(b)^2}{a} + sym \right)
\frac{H(2\gamma)}{H(\gamma)^2}\left(\frac{180}{\pi}\right),
\end{eqnarray}
where $H(\gamma) = \int_{-\infty}^{\infty}dx(1+x^2)^{{-\gamma}/{2}}$.
The second equation is only an approximation, being exact
for $\gamma = 2$. 
Using the above terms, a minimum $\chi^2$ fit to the angular
three-point correlation function 
was performed with the following choice 
\begin{equation}
  \chi^2 = \sum_{a,b,c} \frac{(z(a,b,c) - z_{mod}(a,b,c))^2}{\sigma^2},
\end{equation}
where $z_{mod}$ is equation~\ref{eq:zmod}.
The variance, from Szapudi \& Szalay (1998), is
\begin{equation}
  \sigma^2 = Var(z) = \frac{6 S_3}{S^2\lambda^3} \simeq\frac{6}{DDD}.
\end{equation}
Here $S_3 = \int \Phi(a,b,c)^2$, a configuration integral over 
the definition of the angular bin, described by the function $\Phi(a,b,c) = 1$
when a triangle is in the bin, and zero otherwise.

The simplification arises  since $S_3 = S = \int \Phi$,  
and $S \lambda^3 \simeq DDD$ for our choice of the characteristic
function describing the bin ($\Phi = \Phi^2$)
used in the three-point estimator
(see Szapudi \& Szalay (1998) for details). $DDD$ is the number of
data triplets in the bin.

\subsection{Error Estimation}

The variance defined in the above equation was derived for a Poisson
distribution and therefore accounts only for discreteness effects.
Two additional error contributions, edge and finite
volume effects (Szapudi \& Colombi 1996; Szapudi, Colombi, \& Bernardeau 1999),
arising from the uneven weights
given to data points and from the fluctuations of the universe
on scales larger than the survey size, respectively, are not accounted for.
However, edge effects are expected to be small for these
edge corrected estimators, and it can be shown that 
at least partial correction
for finite volume effects are contained in $DDD$. 
The $\chi^2$'s suggest that this ansatz for the variance is 
within factor of 2 of the truth.  This is the limit of this simple model
ignoring cross correlations of different bins. Without complicated
treatment of the full correlation matrix, it would not make sense 
to improve the above formula with the inclusion of an additional
phenomenological term for finite volume effects to tune
$\chi^2$ to the number of degrees of freedom. While the
results should not sensitively depend on the exact choice
of the error model, it should be kept in mind that our errors 
for $z(a,b,c)$ could be off by as much as a factor of 2.
The most likely sense of the offset is that our errors may be overestimated,
as judged from the $\chi^2$'s.

\clearpage

\hoffset -0.78in
\begin{table}
\begin{center}

{Table 1. $s_3$ and $s_4$ as a function of scale and $I-$band magnitude}
\vskip 20pt

\begin{tabular}{|c|c|c|c|c|c||c|c|c|c|c|}
\hline \hline
 & Scale & & & & & & & & & \\
$I_{min}-I_{max}$ & (degrees) & $s_3$ & $\Delta s_3/s_3$ & $s_4$ & $\Delta s_4/s
_4$ & $I_{min}-I_{max}$ & $s_3$ & $\Delta s_3/s_3$ & $s_4$ & $\Delta s_4/s_4$ \\ \hline
17-18 & 0.01 & 3.54 & 0.53 & 21.80 & L & 
18-19 & 3.48 & 0.10 & \nodata & L \\
$''$ & 0.02 & 5.07 & 0.16 & 75.42 & L & 
$''$        & 3.07 & 0.05 & 2.76 & 0.54 \\
$''$ & 0.04 & 4.64 & 0.10 & 61.34 & L &
$''$        & 3.43 & 0.04 & 12.99 & 0.39 \\
$''$ & 0.08 & 5.04 & 0.11 & 41.44 & L &
$''$        & 4.21 & 0.05 & 24.63 & 0.49 \\
$''$ & 0.16 & 5.46 & 0.19 & 33.71 & L &
$''$        & 4.59 & 0.15 & 41.71 & L \\
$''$ & 0.32 & 3.68 & 0.63 & 12.30 & L &
$''$        & 3.65 & 0.84 & 20.50 & 0.36 \\ \hline
19-20 & 0.01 & 3.40 & 0.04 & 41.57 & 0.49 & 
20-21        & 2.13 & 0.04 & 13.48 & L \\
$''$ & 0.02 & 3.65 & 0.02 & 41.69 & 0.19  &
$''$        & 2.56 & 0.02 & 16.98 & 0.48 \\
$''$ & 0.04 & 4.54 & 0.02 & 53.93 & 0.19 &
$''$        & 2.82 & 0.03 & 17.91 & 0.48 \\
$''$ & 0.08 & 4.89 & 0.05 & 68.70 & 0.36 &
$''$        & 2.29 & 0.09 & 12.15 & 0.44 \\
$''$ & 0.16 & 4.36 & 0.20 & 49.25 & 0.57 &
$''$        & 1.11 & 0.59 & \nodata & L \\
$''$ & 0.32 & 4.62 & L & 2.52 & L  &
$''$        & 1.06 & L & \nodata & L \\ \hline
21-22s  & 0.01 & 2.99 & (0.15,0.13) & 23.64 & (L,0.52)  &
22-22.5s       & 3.56 & (L,0.44) & 129.8 & (L,0.26) \\
$''$ & 0.02 & 2.80 & (0.05,0.05) & 25.98 & (L,0.19) &
$''$        & 2.42 & (0.35,0.21) & 22.41 & (L,0.55) \\
$''$ & 0.04 & 2.15 & (0.08,0.05) & 6.86 & (L,0.49) &
$''$        & 2.38 & (0.31,0.10) & 6.45 & (L,0.58) \\
$''$ & 0.08 & 1.73 & (0.33,0.12) & 5.84 & (L,0.57) &
$''$        & 1.56 & (0.82,0.23) & 20.00 & (L,0.24) \\
$''$ & 0.16 & 0.85 & (L,0.68) & \nodata & (L,L) &
$''$        & 2.70 & (L,0.44) & 50.22 & (L,0.53) \\ \hline 
\end{tabular}
\end{center}
\tablecomments{The relative error $\Delta s_N/s_N$ 
was obtained from the non-linear
FORCE package, which uses a perturbative expansion in terms
of the variances. Thus whenever the result is greater
than $1$, the error-calculation looses accuracy, although
it is a sign that the measurements have low significance
their. For such cases the entry for the table is ``L''.
Systematic error estimates are also provided for the two faintest
slices and are given as the second entry in the parentheses.
The results for slices with $I \le 21$ 
are based on calculations that only exclude data within the
top 5\% largest survey masks. The two faintest slices are based on
calculations in which all masked regions are excluded from analysis.}
\end{table}

\clearpage
\hoffset 0.0in
\begin{table}
\begin{center}

{Table 2. Projection Coefficients: $R_N = S_N / s_N$}
\vskip 20pt

\begin{tabular}{|l|cccc|}
\hline \hline
$I_{min}-I_{max}$ & $R_3$ & $R_4$ & $R_3(\beta = 1)$ & $R_3(\Omega = 1)$\\ 
\hline
17-18 &  1.224 &  1.617 &  1.204  &  1.232 \\
18-19 &  1.074 &  1.212 &  1.189  &  1.062 \\
19-20 &  1.020 &  1.051 &  1.055  &  1.021 \\
20-21 &  1.049 &  1.134 &  1.024  &  1.053 \\
21-22 &  1.069 &  1.189 &  1.051  &  1.071 \\
22-22.5 &  1.076 &  1.215 &  1.06.5 &  1.07750 \\
\hline
\end{tabular}
\end{center}

\tablecomments{ The columns $R_3(\beta = 1)$ and $R_3(\Omega = 1)$
illustrate how much $R_3$ changes if $\beta = 1$ (the luminosity function
evolution parameter -- see \S\ref{sec:cic}) or $\Omega = 1$ is used.}
\end{table}
 
\clearpage
\begin{table}
\begin{center}

{Table 3. $\chi^2$ fits for the general connected third order
expansion of the three-point function.}
\vskip 20pt

\begin{tabular}{|c|c|c|c|c|c||c|c|c|c|}
\hline \hline
                &         & \multicolumn{4}{c||}{Fits with $0.1 < q_3 < 10$} 
                          & \multicolumn{4}{c|}{Fits with $0.01 < q_3 < 10$} \\
\hline
$I_{min}-I_{max}$ & $N_f$ & $\chi^2$ & $q^3$ & $q^{111}$ & $q^{21}$ 
                          & $\chi^2$ & $q^3$ & $q^{111}$ & $q^{21}$ \\ \hline
17 - 18   & 44 & 11.28 & 1.29 & -0.26 & 0.03 &
                 11.38 & 1.30 & -0.25 & 0.03 \\
$''$      & 42 & 12.24 & 1.76 & & &
                 12.30 & 1.76 & & \\ \hline
18 - 19   & 52 & 15.03 & 0.86 & 0.48 & 0.01 &
                 15.03 & 0.86 & 0.48 & 0.01\\
$''$      & 50 & 22.98 & 1.39 & & &
                 22.98 & 1.39 & & \\ \hline
19 - 20   & 62 & 81.92 & 3.18 & 0.29 & -0.07 &
                 88.90 & 3.24 & 0.75 & -0.11\\
$''$      & 60 & 85.69 & 2.80 & & &
                 96.45 & 2.76 & & \\ \hline
20 - 21   & 41 & 100.31 & 0.91 & -2.35 & 0.16 &
                 107.64 & 1.02 & -2.11 & 0.10\\
$''$      & 39 & 116.11 & 1.00 & & &
                 126.28 & 0.86 & & \\ \hline \hline
21 - 22   & 17 & 10.92 & 0.29 & -0.27 & 0.01 &
                 12.96 & 0.15 & -0.38 & 0.03\\
$''$      & 15 & 11.66 & 0.27 & & &
                 14.06 & 0.18 & & \\ \hline
22 - 22.5 & 29 & 28.32 & 0.45 & -0.57 & 0.02 &
                 36.22 & 0.47 & -0.47 & 0.00\\
$''$      & 27 & 34.15 & 0.43 & & &
                 45.38 & 0.32 & & \\ \hline \hline
21 - 22s  & 19 &  9.95 & 0.51 & -0.80 & -0.00 &
                 13.07 & 0.32 & -0.81 & 0.03\\
$''$      & 17 & 12.30 & 0.34 & & &
                 15.20 & 0.24 & & \\ \hline
22 - 22.5s& 42 & 25.32 & 0.48 & -1.69 & 0.13 &
                 29.96 & 0.39 & -1.75 & 0.14\\
$''$      & 40 & 33.92 & 0.57 & & &
                 39.56 & 0.52 & & \\ \hline
\hline
\end{tabular}
\end{center}

\tablecomments{Results for the deepest slices,
$I \ge 21$, are shown with 
(mag limits followed by ``s'') and without 
statistical corrections for 
star/galaxy misclassification (see \S\ref{sec:3pcf}).}
\end{table}

%% FIGURES & CAPTIONS HERE:
\clearpage

\end{document}